\documentclass[12pt,twocolumn,usenatbib]{mnras}
\usepackage[T1]{fontenc}
\usepackage{ae,aecompl}
\usepackage{graphicx}	
\usepackage{amsmath}	
\usepackage{amssymb}	
\usepackage{natbib}

\title[Do the planets exist?]{Do the planets in the HD 34445 system really exist?}
\author[Nikolaos Georgakarakos and Ian Dobbs-Dixon]{
Nikolaos Georgakarakos,$^{1}$\thanks{E-mail: georgakarakos@hotmail.com}
and Ian Dobbs-Dixon$^1$ \\
$^{1}$New York University Abu Dhabi, Saadiyat Island, P.O. Box 129188, Abu Dhabi, UAE}
\date{Accepted XXX. Received YYY; in original form ZZZ}

\pubyear{2019}
\begin{document}
\label{firstpage}
\pagerange{\pageref{firstpage}--\pageref{lastpage}}
\maketitle

\begin{abstract}
In 2010 the first planet was discovered around star HD 34445. Recently,  another five planets were announced orbiting the same star. It is a rather dense multi-planet system with some of its planets having separations of fractions of an au and minimum masses ranging from Neptune to sub-Jupiter ones. Given the number of planets  and the various uncertainties in their masses and orbital elements,  the HD 34445 planetary system is quite interesting as there is the  potential for mean motion and secular resonances that could render the outcome of its dynamical evolution and fate an open question.  In this paper we investigate the dynamical stability of the six planet system in order to check the validity of  the orbital solution acquired.
This is achieved by a series of numerical experiments,  where the dynamical evolution of the system is tested on different timescales. We vary the orbital elements and masses of the system within the error ranges provided. We find that for a large area of the parameter space  we can produce stable configurations and therefore conclude it is very likely that the HD 34445 planetary system is real.  Some discussion about the potential habitability of the system is also done.
\end{abstract}



\section{Introduction}
The number of confirmed exoplanets rises day after day, and  almost half of them reside in systems with more than 
one planet.  Only a handful of those systems, however, have six or more planets.  The most populous system so far is Kepler-90 \citep{2014ApJ...781...18C,2018AJ....155...94S} having eight planets, followed by the TRAPPIST-1 system that contains seven planets  \citep{2016Natur.533..221G}.  Each of Kepler-11 \citep{2011Natur.470...53L},  Kepler-20 \citep{2012ApJ...749...15G,2012Natur.482..195F,2016AJ....152..160B},  Kepler-80 \citep{2013ApJS..208...22X,2014ApJ...784...45R,2016ApJ...822...86M,2018AJ....155...94S},  HD 10180 \citep{2011A&A...528A.112L} and HD 219134 \citep{2015A&A...584A..72M,2015ApJ...814...12V} have six planets.

\begin{figure*}
\begin{center}
\includegraphics[width=180mm,height=20mm]{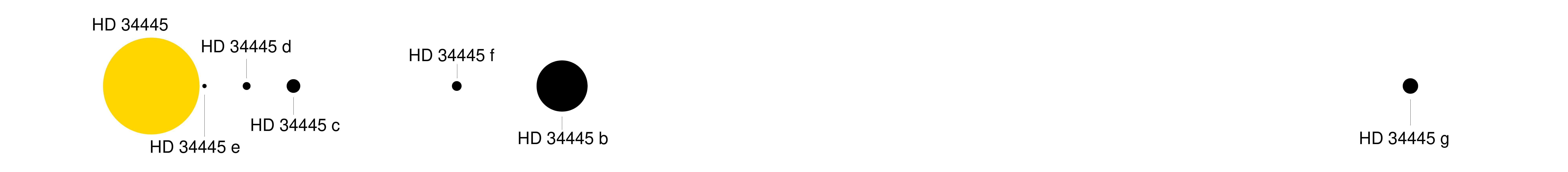}
\caption{A schematic representation of the HD 34445 system.  The distances between the dots and their relative sizes (excluding the star) reflect the distances and masses of the real system.
\label{fig1}}
\end{center}
\end{figure*}

Recently, the HD 34445 system joined the six planet system family.  The first planet of the system, HD 34445 b, was discovered by the California Planet survey  in 2010 \citep{2010ApJ...721.1467H}.  They announced a planet with $M\sin{i} = 0.79 M _{Jup}$ ($M$ being the mass of the planet and $i$ the orbital inclination with respect to the plane of the sky) in
a mildly eccentric $(e = 0.27)$ orbit with a 2.87 year period.  Their single planet fit, however,  had relatively large residuals, implying
the presence of more planets in the system.  \cite{2010ApJ...721.1467H} stated that the underestimated stellar jitter could explain the large residuals, but they thought that was less likely given the metallicity, colour, and modest evolution of HD 34445 which is an old G0V dwarf star.
The former explanation was recently verified  when \cite{2017AJ....154..181V} announced the discovery of another five planets around HD 34445.  The planets are on nearly circular orbits with their  masses being several times the mass of the Earth.  The eccentricity of the first planet discovered in the system, HD 34445 b, was also revised, placing that planet on an almost circular orbit too. The most distant planet of the system, HD 34445 g, however, has a rather long period of around 5700 days.  That means more RV measurements will be required in order to better
constrain the parameters of HD 34445 g.  Figure \ref{fig1} is a graphical representation of the HD 34445 planetary system.

An issue of great interest but also a useful tool for validating the orbital solution and placing constrains to the planetary orbits is the dynamical stability of multi-planet systems.  This kind of problem, however, is rather complex due to the number of bodies involved and the uncertainties that come with the orbital elements and physical parameters of the system under investigation.  As a result, quite often, the relevant investigation does not reach a convincing conclusion and the question may remain open. 

Regarding Kepler-90, there has not been a study about the stability of all eight planets. 
Some work was done in the discovery paper of \cite{2014ApJ...781...18C} by means of numerical simulations. Although not described in much detail, it seems that a series of numerical experiments was carried out and the seven planet system seemed to be close to instability (the authors actually ignored the two innermost planets as they did not see any sign of dynamical interactions with the rest of the planets). As part of their work of exploring the dynamical stability of tightly packed planetary systems under the perturbations of an additional planet, \cite{2018AJ....155..139G} also investigated the stability of Kepler-90. By performing N-body simulations they found that $20\%$ of the orbital configurations they used got unstable.
The total number of configurations they simulated, however, was only 100.

The dynamical stability of the TRAPPIST-1 system has also been under investigation. In the  announcement paper of the seven planet system, \cite{2017Natur.542..456G} used N-body simulations to assess the evolution of a number of orbital configurations over a time span of 0.5 Myrs. Most of the integrations ended up with the disruption of the system.  The authors also used a statistical method to assess the stability of the TRAPPIST-1 system
and they found that the system had $25\%$ chance to suffer an instability over 1 Myrs.  The inclusion of tidal effects in simulating the system seemed to postpone the disruption of the system in most cases. Their conclusion was that the stability of the system 
depended a lot on the values of the masses and of the orbital elements of the planets, as well as on the inclusion or exclusion of planet h.  Additional simulations were performed by \cite{2017ApJ...842L...5Q} in order to place constrains to the system's parameters and 
try to identify plausible compositions for the planets.  Out of the 18527 systems they numerically integrated, $28.2\%$ proved to be stable over 1 Myrs.  Things seem to become more favourable for the system's long term stability when disk migration is taken into consideration. \cite{2017ApJ...840L..19T} managed to produce stable configurations of the
TRAPPIST-1 system that were initialised through disk migration. Those stable configurations were stable for at least 50 Myrs, even without tidal dissipation.  

Similar stability studies have been done for the six planet systems.  In the discovery article of Kepler-11, \cite{2011Natur.470...53L} integrated several systems that fitted their transit data for 250 Myrs. They found that long term stability of the system is possible, although weak chaos was evident in the mean motions and eccentricities of the planets. Through a more extensive dynamical analysis of the system, \cite{2012MNRAS.427..770M} found that the 
dynamics of the system are determined by three and four body mean motion resonances. The  overlapping of those resonances results in unstable orbital configurations. In a later study, \cite{2017MNRAS.470.4145H} used a Monte-Carlo method to generate initial conditions and they carried out N-body integrations of Kepler-11. The authors of that work found that most of their simulated systems were
unstable, resulting in orbit crossings, leading eventually to collisions and mergers. 

Contrary to Kepler-11, Kepler-20 seems to be more stable in the long term.  By running numerical simulations of the five planet Kepler-20 system (Kepler-20 g was discovered a few years later), \cite{2012ApJ...749...15G} concluded that low eccentricity models were stable over long timescales. This dynamical picture was repeated even with the inclusion of Kepler-20 g \citep{2016AJ....152..160B}.  

Similarly, \cite{2016AJ....152..105M}, based on analytical and empirical stability formulae, analysed the Kepler-80 system consisting of five planets (Kepler-80 g was confirmed a couple of years later) and they concluded that the system was stable as long as the eccentricities were below $\sim 0.2$.  

The HD 10180 system was also found to be stable. \cite{2011A&A...528A.112L} numerically integrated the planetary orbits over 150 Myrs and they also carried out a global frequency analysis \citep{1993PhyD...67..257L}. Their investigation, however, was limited to the nominal solution only. That result was in agreement with \cite{2012A&A...543A..52T}, where the hypothesis of the existence of additional planets in the HD 10180 system was investigated.  Following \cite{2006ApJ...647L.163B}, \cite{2012A&A...543A..52T} used a Lagrange stability criterion on the six planet model of HD 10180 and he concluded that his result was in line with the numerical experiment outcome of \cite{2011A&A...528A.112L}.

Finally, regarding the HD 219134 system, \cite{2015A&A...584A..72M} performed some simulations of the four planet version of the system. They found that the system was stable over $10^6$ orbits of the outermost planet.

The dynamical stability of the nominal solution of the six planet HD 34445 system was briefly investigated in \cite{2017AJ....154..181V}.
Using a Bulirsch-Stoer integrator, the nominal solution of the system initialized with zero eccentricities (private communication S. Vogt) was simulated for 10000 years and it was found to be stable. In this work we investigate the 
dynamical stability of the HD 34445 system in more detail. Using a series of numerical experiments and taking into consideration the uncertainties in the physical parameters and orbital elements of the planets we aim to see whether the six planet configuration is robust and try to place stricter constraints on the system parameters.

\section{Method}
\label{met}
In order to investigate the dynamical stability of HD 34445,  we carry out a series of numerical experiments.  As 
stability can have many definitions \citep[e.g.][]{1984CeMec..34...49S,2008CeMDA.100..151G}, in the context of this work 
the system is classified stable if no planetary orbit becomes hyperbolic with respect to the star  at any point during the integration time and no orbit crossing takes place between any of the planets  at any moment.   We also assume that all bodies lie on the same plane of motion. 
For each planet we have its  minimum mass and four orbital elements, i.e. the semi-major axis, the eccentricity, the argument of pericentre and the mean anomaly. Each of these quantities is accompanied by its $1\sigma$ error and a $99\%$ credibility interval given by \cite{2017AJ....154..181V} . These are the parameter ranges over which we explore the dynamical stability of the system.  All the orbital elements and physical parameters along with their uncertainties can be found in Table 1 (as given in Table 9 of \cite{2017AJ....154..181V}).  

\begin{figure}
\begin{center}
\includegraphics[width=85mm,height=60mm]{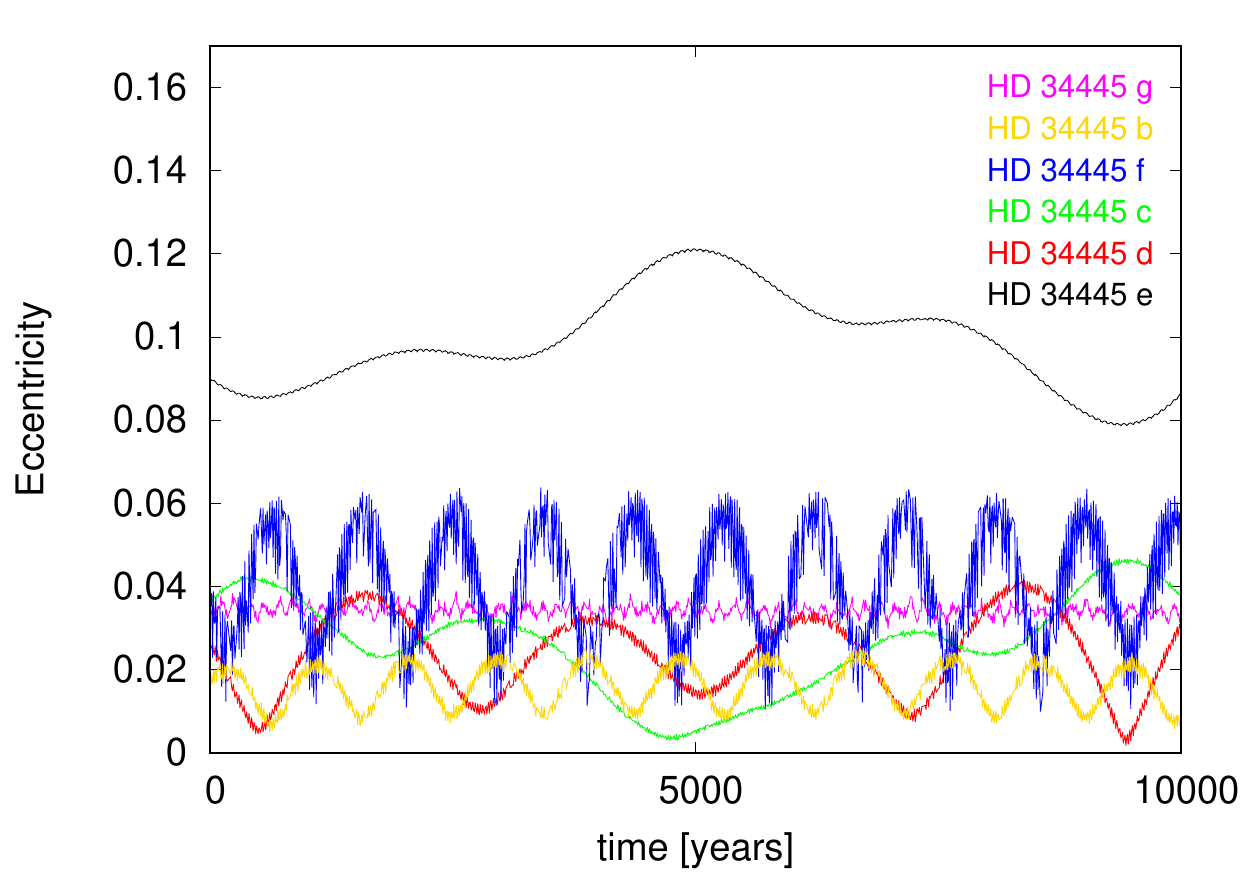}
\caption{Eccentricity evolution for all the HD 34445 planets over ten thousand years.
\label{fig2}}
\end{center}
\end{figure}

The stability of the system is evaluated through three different numerical experiments.  In the first experiment,
we integrate forward in time the nominal solution for five million years.  We remind the reader that in \cite{2017AJ....154..181V}, the nominal solution was simulated for only ten thousand years. Although this amount of time covers the longest of the secular periods of the planets in the system, it is likely that some instabilities may take longer time to demonstrate. Therefore we wanted to make sure that we do not miss any important dynamical events.  Figure \ref{fig2} shows the eccentricity evolution of all the planets of the system over 10000 years so that the reader can note the different periods.

In the second experiment we vary one of the five parameters mentioned above, i.e. planetary mass, semi-major axis, eccentricity, argument of pericentre and mean anomaly.  Each time, one of these quantities is allowed to vary for all the planets while for the rest of the quantities we use the values of the nominal solution. Each varied parameter gets three values:
the nominal one plus minus the minimum and maximum uncertainties for the $1\sigma$ case or the endpoints of the  $99\%$ credibility interval. Hence,
for each one of those single parameter cases we simulate $3^6=729$  systems.  An exception to testing the parameters within both the $1\sigma$  and $99\%$ credibility interval limits  is the mean anomaly and argument of pericentre.  As the credibility interval for both the mean anomaly and argument of pericentre is $[0,2\pi]$, we do not run any simulations varying the two parameters over that range and
we only make use of their  $1\sigma$  range.  The simulation time for all the simulations of the second experiment is set to one million years initially. In case we find unstable systems for any of the single parameter cases, we increase the simulation time to five million years and we rerun all the 729 combinations for that specific parameter.

Finally,  in the third experiment, we vary the orbital inclination with respect to the plane of the sky.  Since the planetary system was discovered by the radial velocity method, we only know the minimum masses of the planets.  It is quite likely that the real masses are larger. Therefore it is sensible to test the limit of the stability of the system as a function of the masses. 
Hence, we integrate the nominal solution of the system but with orbital inclinations ranging from $i=1^{\circ}$ to $i=89^{\circ}$ with a step of one degree. The simulations are carried out for 5 million years.  The planets are kept coplanar as the line of sight is varied.

For the simulations we used the Gauss-Radau integrator in \cite{2010LNP...790..431E}. We have set the code to stop the integration when any planet becomes hyperbolic.

\begin{table*}
\label{t1}
\caption[]{Masses and orbital elements for the HD 34445 system. The values are given with both standard $1\sigma$ errors and $99\%$ credibility intervals that arise from the statistical model used for the RV data fitting. The stellar mass is $1.07\pm 0.02 M_{\odot}$ and its uncertainty has been taken into consideration in the calculation of the semi-major axes and minimum masses.} 
\begin{scriptsize}
\vspace{0.1 cm}
\begin{center}	
{\footnotesize \begin{tabular}{c c c c c c}\hline
Body & Mass & Semi-major axis& Eccentricity & Mean Anomaly & Arg. of pericentre \\
             &  (in $M_{\oplus}$)& (in au)&  & (in radians)& (in radians)\\
\hline
HD 34445 e & $16.8\pm 2.9 \hspace{0.1cm}[8.1, 24.8]$ & $0.2687\pm 0.0019\hspace{0.1cm}[0.2632, 0.2741]$ & $0.090\pm 0.062 \hspace{0.1cm}[0, 0.287]$ & $0.7\pm 1.8 \hspace{0.1cm}[0, 2\pi]$ & $5.3\pm 2.1 \hspace{0.1cm}[0, 2\pi]$\\ 
HD 34445 d & $30.7\pm 3.9 \hspace{0.1cm}[19.2, 41.2]$ & $0.4817\pm 0.0033\hspace{0.1cm}[0.4715, 0.4909]$ & $0.027\pm 0.051 \hspace{0.1cm}[0, 0.234]$ & $2.2\pm 1.6 \hspace{0.1cm}[0, 2\pi]$ & $4.3\pm 1.7 \hspace{0.1cm}[0, 2\pi]$\\ 
HD 34445 c & $53.5\pm 5.0 \hspace{0.1cm}[40.0, 68.4]$ & $0.7181\pm 0.0049\hspace{0.1cm}[0.7034, 0.7314]$ & $0.036\pm 0.071 \hspace{0.1cm}[0, 0.217]$ & $4.1\pm 1.4 \hspace{0.1cm}[0, 2\pi]$ & $2.6\pm 1.3 \hspace{0.1cm}[0, 2\pi]$\\ 
HD 34445 f & $37.9\pm 6.5 \hspace{0.1cm}[18.7, 57.0]$ & $1.543\pm 0.016\hspace{0.1cm}[1.500, 1.590]$ & $0.031\pm 0.057 \hspace{0.1cm}[0, 0.264]$ & $1.4\pm 1.8 \hspace{0.1cm}[0, 2\pi]$ & $3.7\pm 1.5 \hspace{0.1cm}[0, 2\pi]$\\ 
HD 34445 b & $200.0\pm 9.0 \hspace{0.1cm}[175.0, 224.9]$ & $2.075\pm 0.47\hspace{0.1cm}[2.034, 2.120]$ & $0.014\pm 0.035 \hspace{0.1cm}[0, 0.117]$ & $3.9\pm 1.6 \hspace{0.1cm}[0, 2\pi]$ & $2.0\pm 1.6 \hspace{0.1cm}[0, 2\pi]$\\ 
HD 34445 g & $120.6\pm 38.9 \hspace{0.1cm}[48.7, 257.8]$ & $6.36\pm 1.02\hspace{0.1cm}[5.04, 10.29]$ & $0.032\pm 0.080 \hspace{0.1cm}[0, 0.259]$ & $3.0\pm 1.6 \hspace{0.1cm}[0, 2\pi]$ & $4.1\pm 1.6 \hspace{0.1cm}[0, 2\pi]$\\ 
\hline
\end{tabular}}
\end{center}
\end{scriptsize}
\end{table*}

\section{Results}
In this section we present the results of each one of the three numerical experiments we have carried out.

\subsection{Stability of the nominal solution}
\begin{figure}
\begin{center}
\includegraphics[width=80mm,height=55mm]{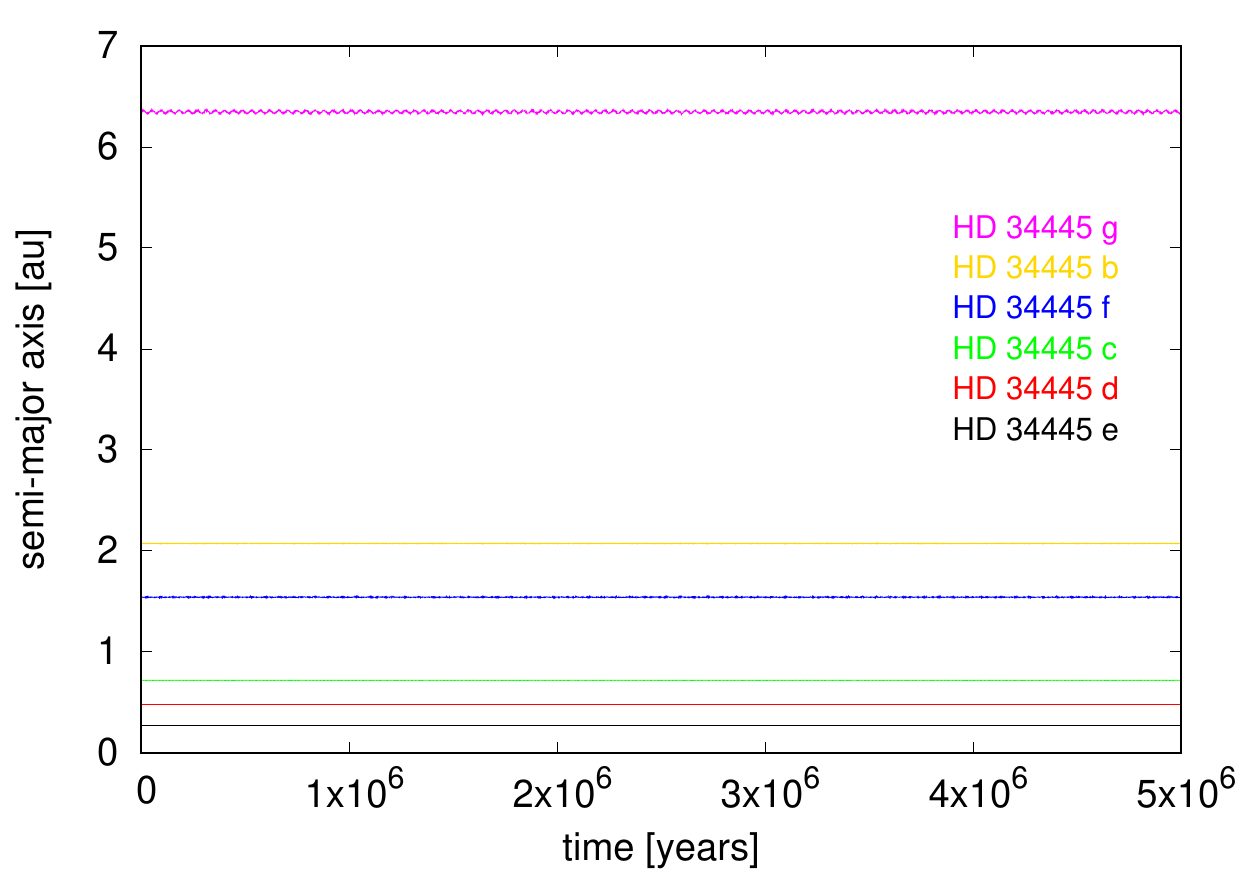}
\includegraphics[width=80mm,height=55mm]{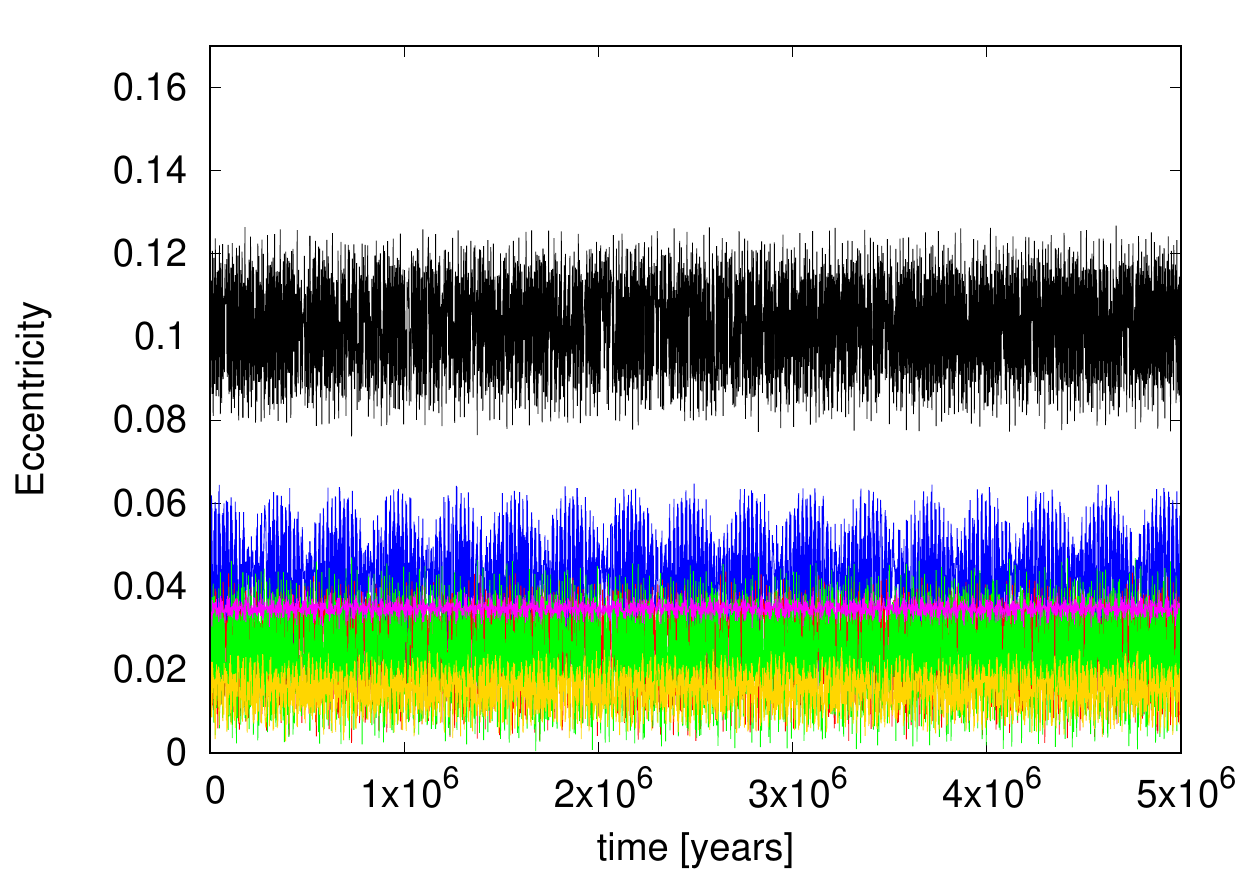}
\caption{Semi-major axis and eccentricity evolution.   The nominal orbital solution of the HD 34445 system integrated forward in time for 5 million years.
\label{fig3}}
\end{center}
\end{figure}

The five million year numerical integration of the nominal solution of the HD 34445 system
showed neither orbit crossing between any of the planets nor any planetary orbit becoming hyperbolic with respect to the star.  The 
planetary semi-major axes showed very little fluctuations over orbital timescales (the greatest deviation from a nominal value was around $0.5\%$), while the eccentricity
evolution did not reveal any dramatic increase for any of the planets.  The maximum eccentricity value achieved was 
around 0.12 for HD 34445 e while for the rest of the planets the maximum eccentricity remained under 0.07.  Figure \ref{fig3} exhibits the semi-major axis and eccentricity evolution of all six planets.

\subsection{One parameter variation experiment}
In the second numerical experiment the results showed a mixed behaviour. 
Varying any of the semi-major axis, minimum mass, mean anomaly or argument of  pericentre
either at the $1\sigma$ or at the $99\%$ credibility interval level produced no unstable systems for the 
integration time chosen. 

On the other hand, varying the eccentricity, led in many
cases to instability. Therefore we extended our simulations to five million years.
In the $1\sigma$ case, out of the 729 systems integrated, 135 had at least one planet becoming hyperbolic, while 6 systems had orbit crossing incidents between at least two planets but without any hyperbolic orbits recorded.  It is quite likely that those kind of systems will eventually end up with some planet going hyperbolic if we extend the integration time.  The smallest planet of the system was the one with the most hyperbolic orbit incidents recorded. HD 34445 e became hyperbolic in 82 cases, followed by the second smallest planet, HD 3444 d, with 31 times.  HD 34445 c became hyperbolic 3 times, while HD 34445 f became hyperbolic only once. Finally, the two outermost and largest planets of the system remained on elliptic orbits in all cases. Most of the times a planetary eccentricity became greater than 1, only one planet was involved (111 cases). There were, however, cases where two planets (20 times), three planets (3 times) and even four planets (1 time) became hyperbolic.  

Additionally, for the orbit crossing and hyperbolic orbit cases we counted how many times each one of the three values (minimum,nominal and maximum) was involved. For example, in every single case we had an unstable system, the maximum value of the eccentricity of HD 34445 b was involved in 93 cases, while the nominal value was involved in 48 cases. For planet HD 34445 f, the maximum eccentricity value was involved 128 times while the minimum value was involved in 12 cases. In only 1 case the nominal value was present. For the rest of the planets the value distribution for the unstable cases was more balanced.  

Finally, we had a look at the effect of the integration time on the number of unstable cases. The majority of the systems became unstable in less than one million years. 95 out of our 141 cases exhibited orbit crossing incidents or hyperbolic orbits within that time span. In the next million years we had another 19 cases making the total number to be 114. That number rose to 127 in the next million years, while the total number rose to 134 by the end of four million years. The last  7 cases were recorded between  four and five million years.
Hence, it looks that longer integration times will probably add a few more systems to our numbers but will not affect the qualitative picture for the $1\sigma$ eccentricity case.

\begin{figure*}
\begin{center}
\includegraphics[width=75mm,height=70mm]{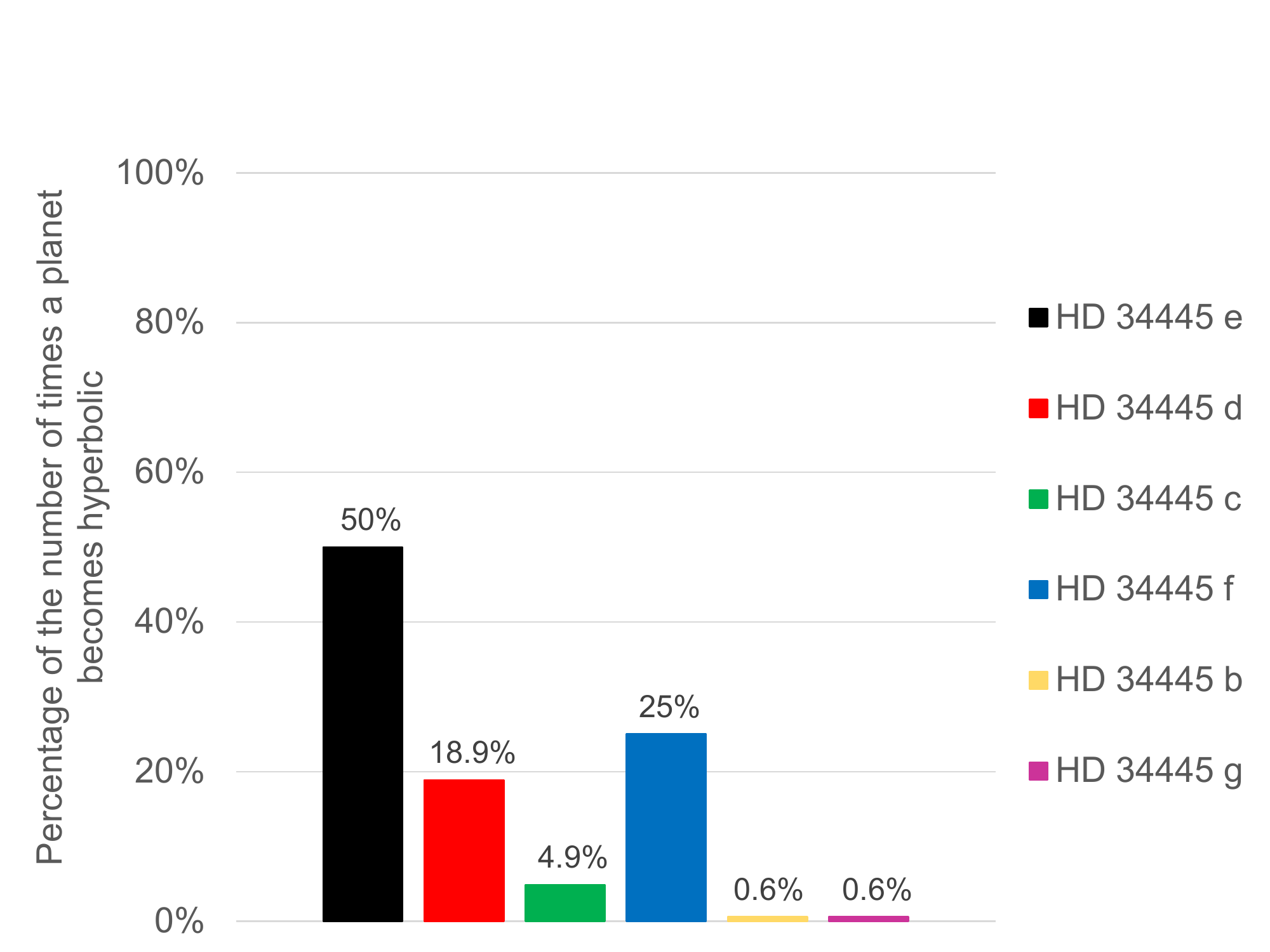}
\includegraphics[width=90mm,height=70mm]{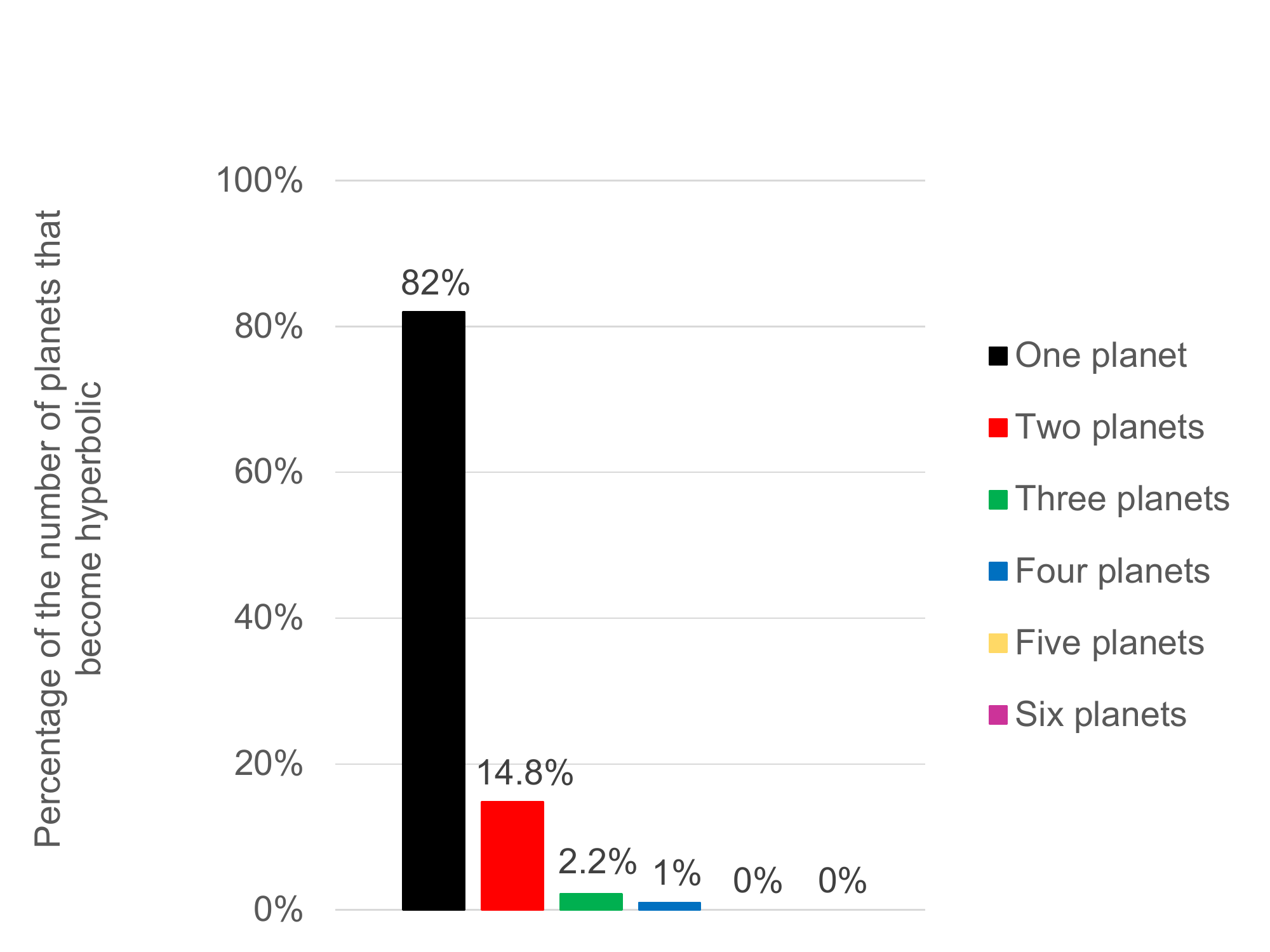}
\includegraphics[width=80mm,height=60mm]{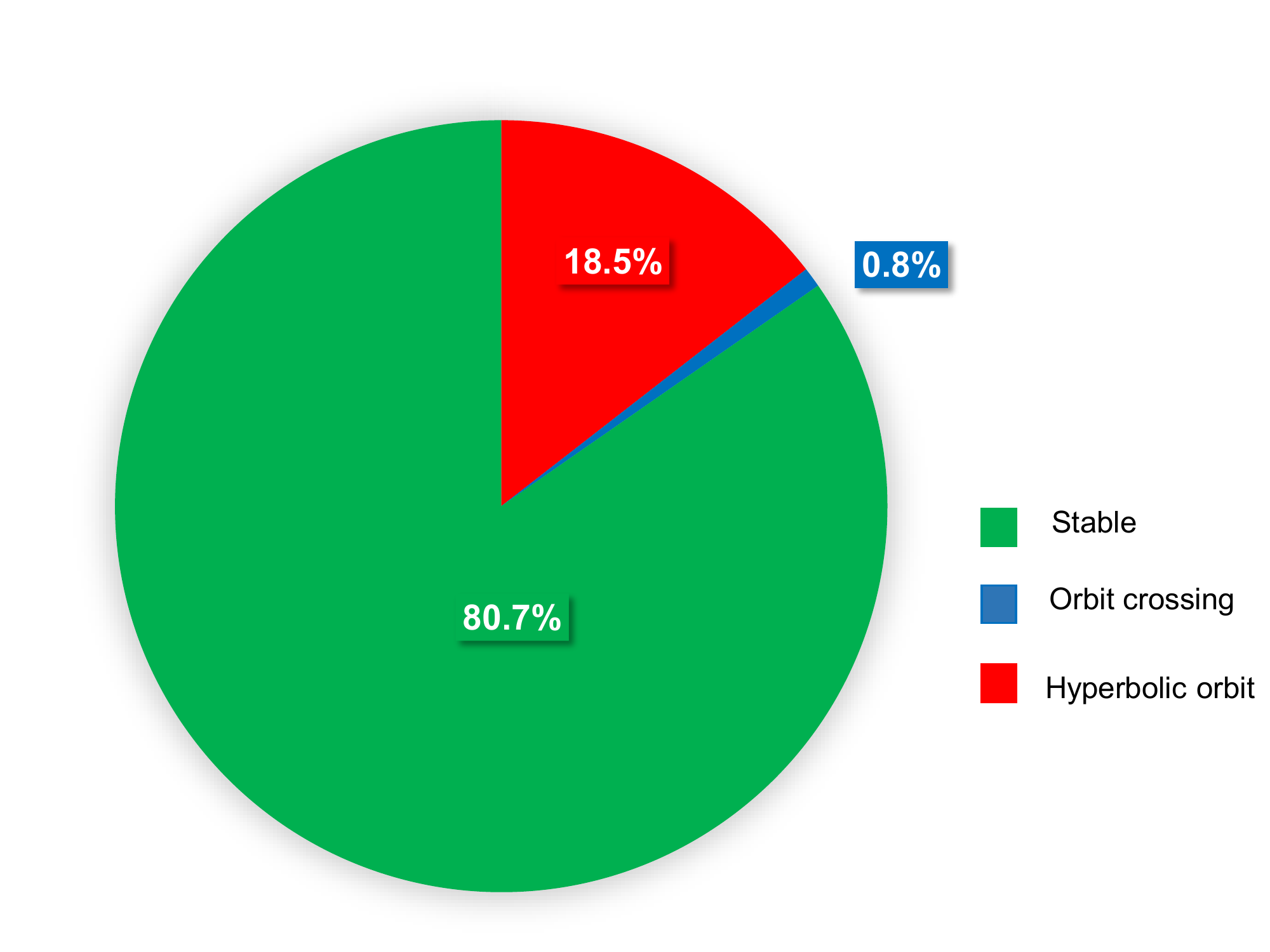}\\
\includegraphics[width=90mm,height=70mm]{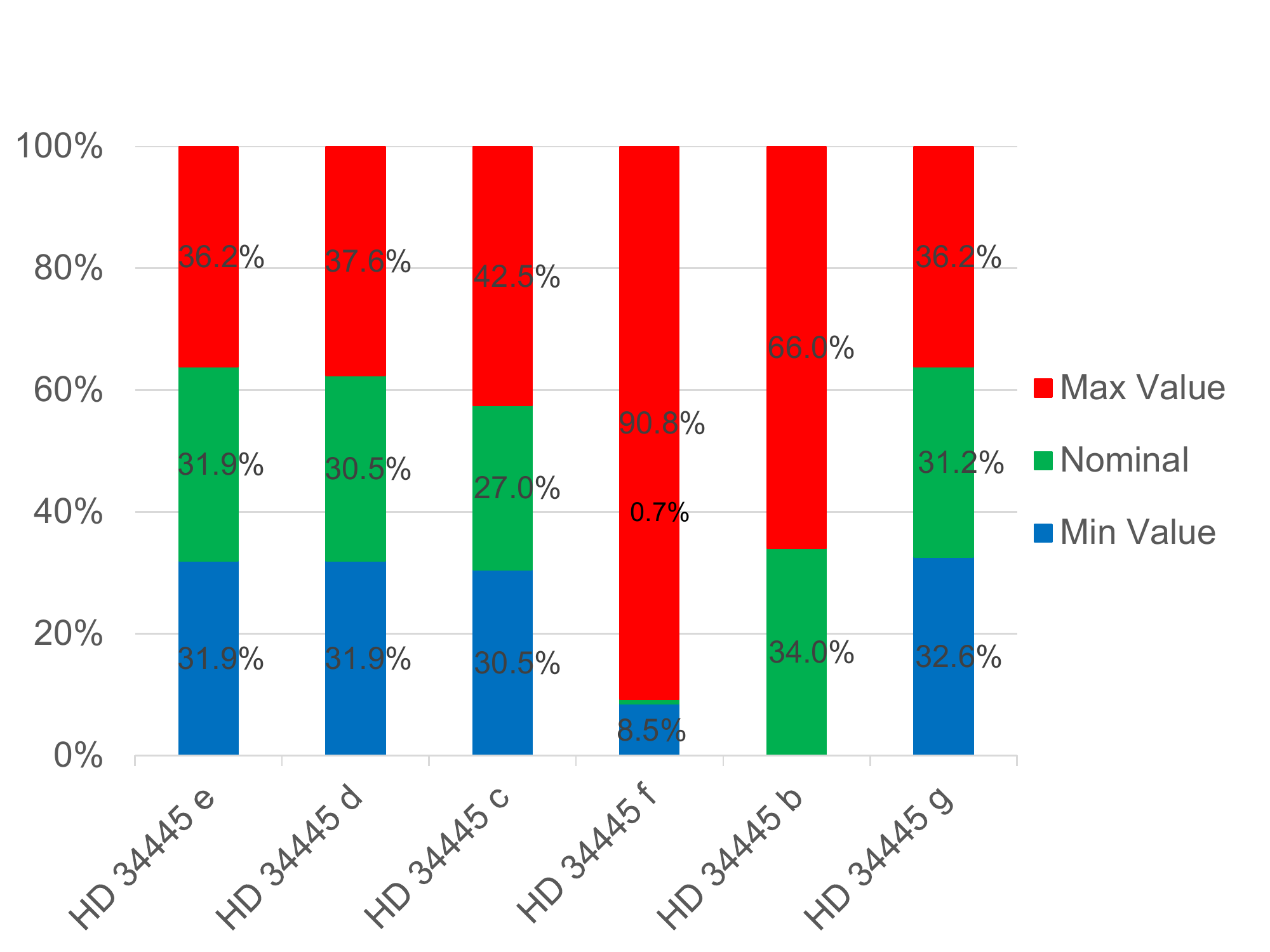}
\includegraphics[width=70mm,height=60mm]{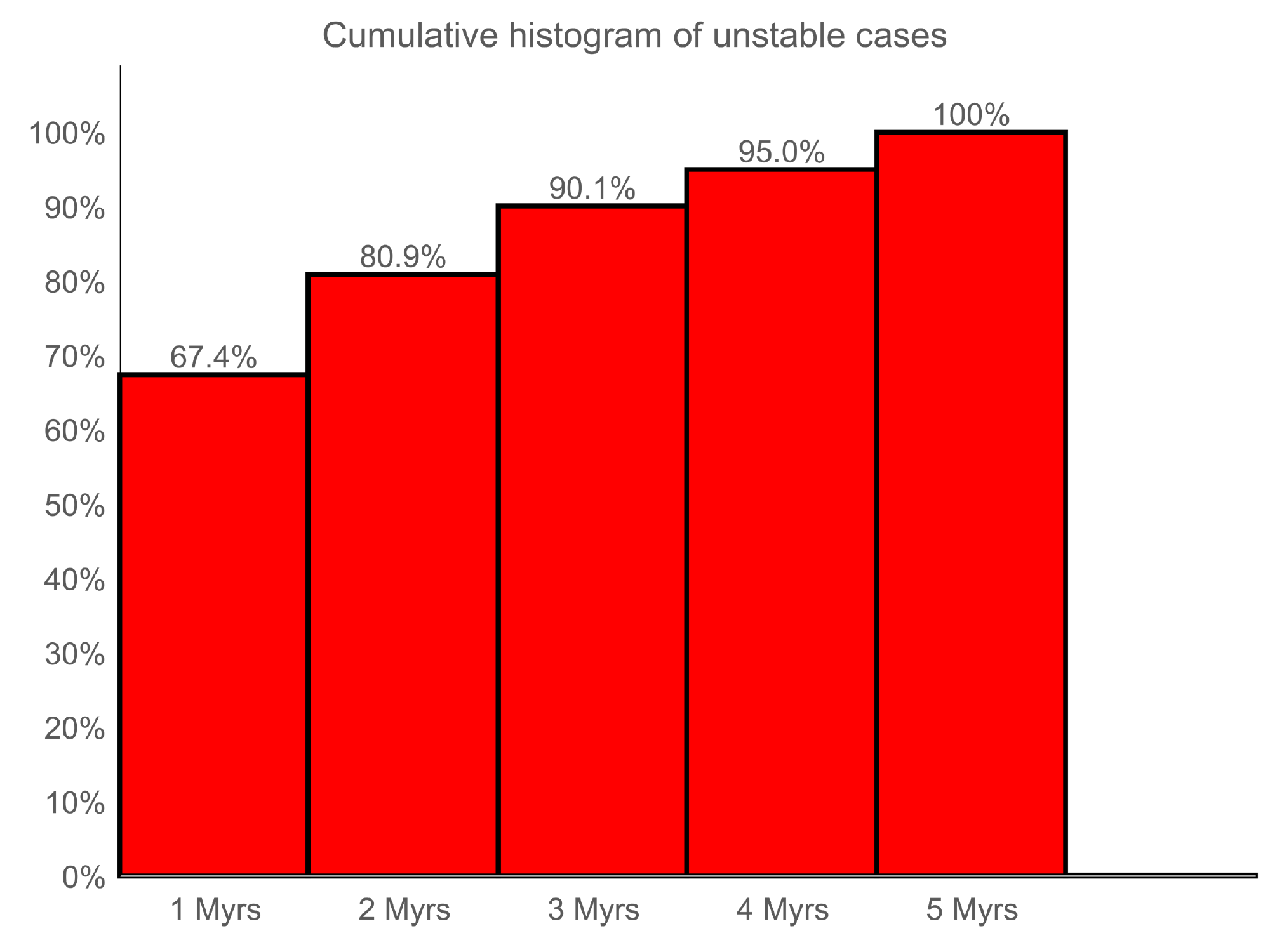}
\caption{The $1\sigma$ eccentricity case integrated for five million years. Top left: percentage of hyperbolic orbits for each planet. Top right:  percentage of  planets that became hyperbolic during a simulation. Middle: percentage of stable systems, hyperbolic orbits and orbit crossing incidents. Bottom left: distribution of which of the three values (minimum, nominal, maximum) used to sample our parameters was involved in the unstable systems recorded (hyperbolic + orbit crossing cases). Bottom right: cumulative histogram of the percentage of unstable
systems with respect to the time of integration. We recorded 135 hyperbolic orbit and 6 orbit crossing cases.
\label{fig4}}
\end{center}
\end{figure*}

Figure \ref{fig4} is a graphical representation of the results for the $1\sigma$ eccentricity  case.

\begin{figure*}
\begin{center}
\includegraphics[width=75mm,height=70mm]{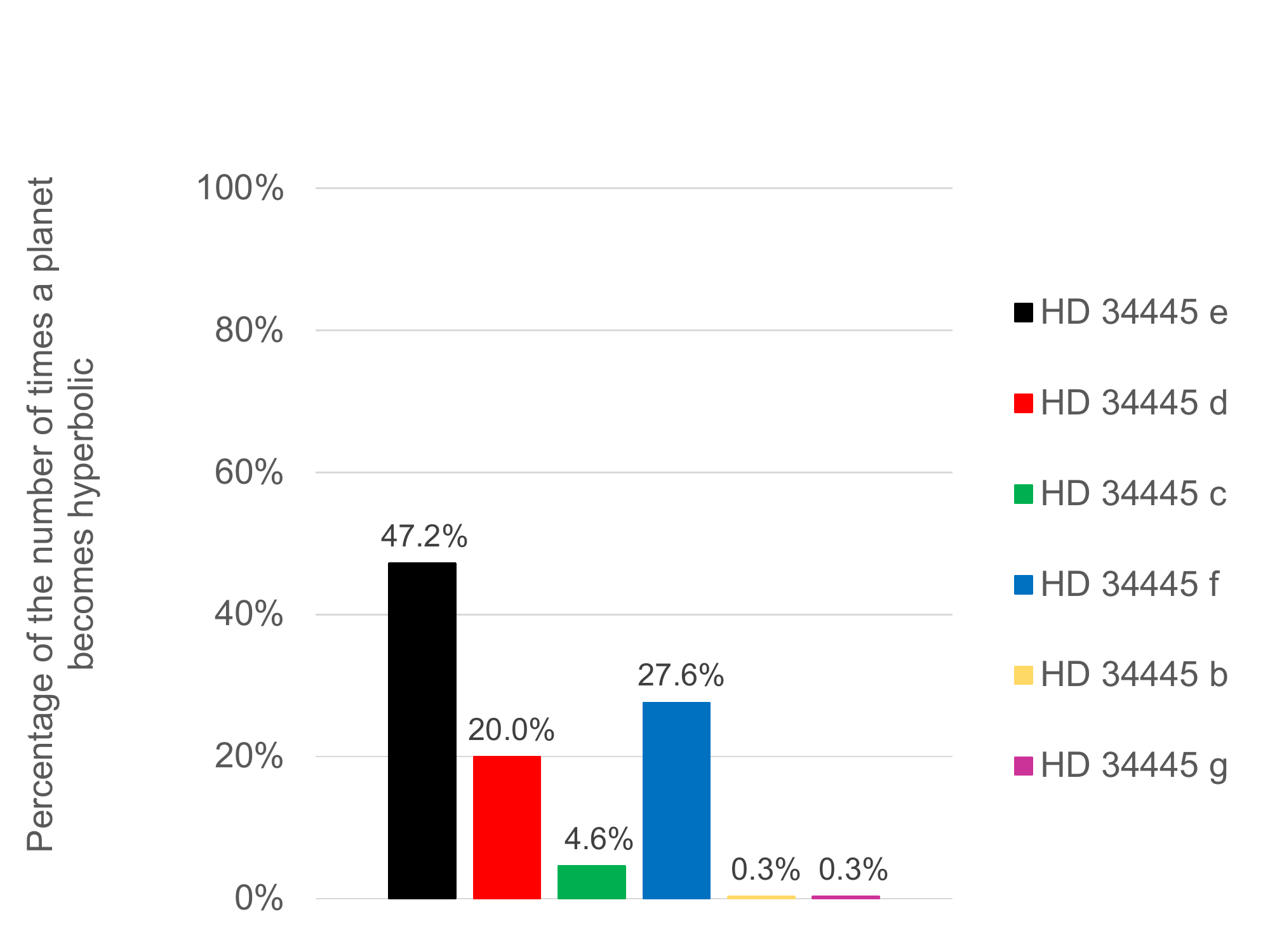}
\includegraphics[width=90mm,height=70mm]{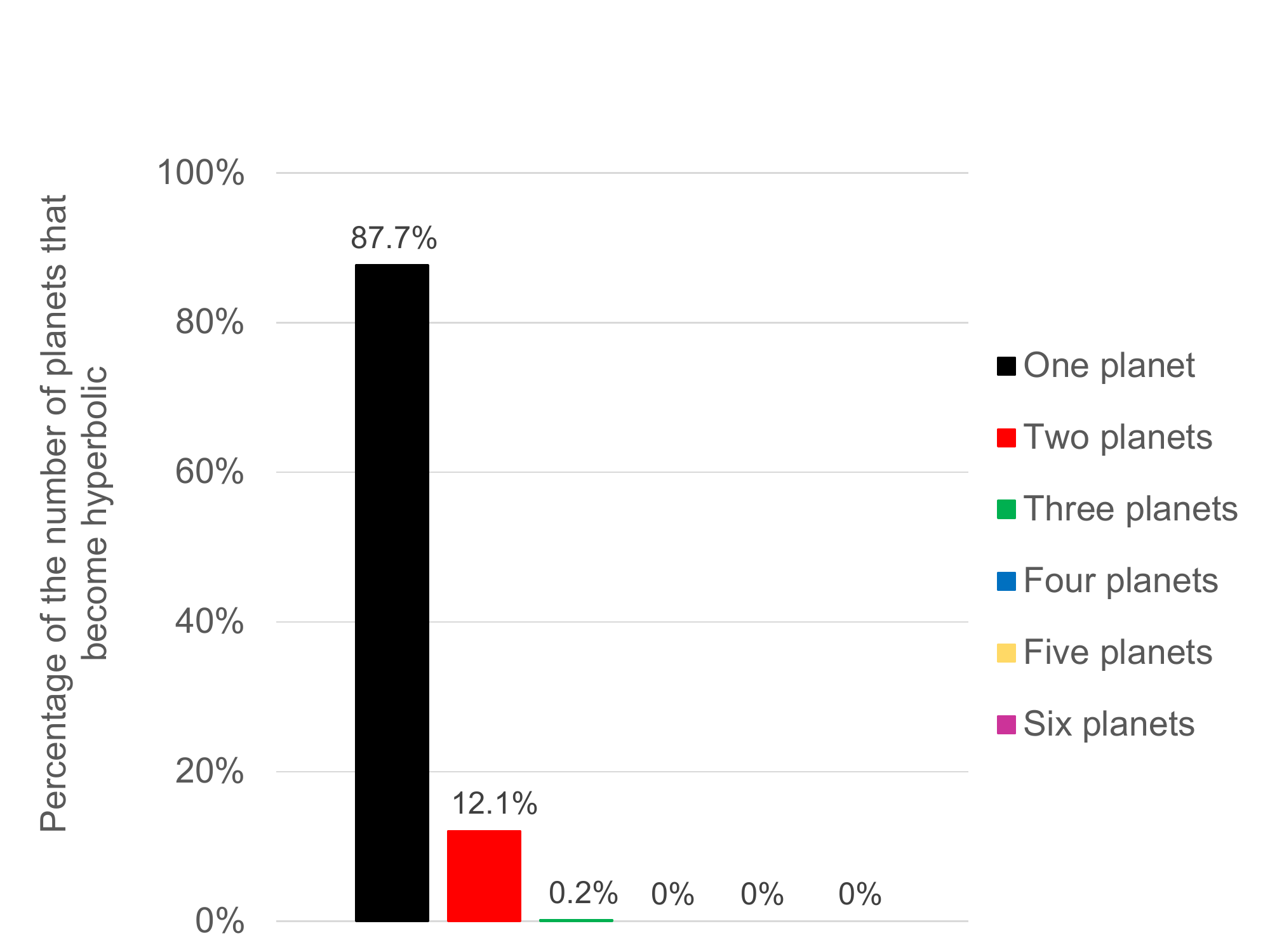}
\includegraphics[width=80mm,height=60mm]{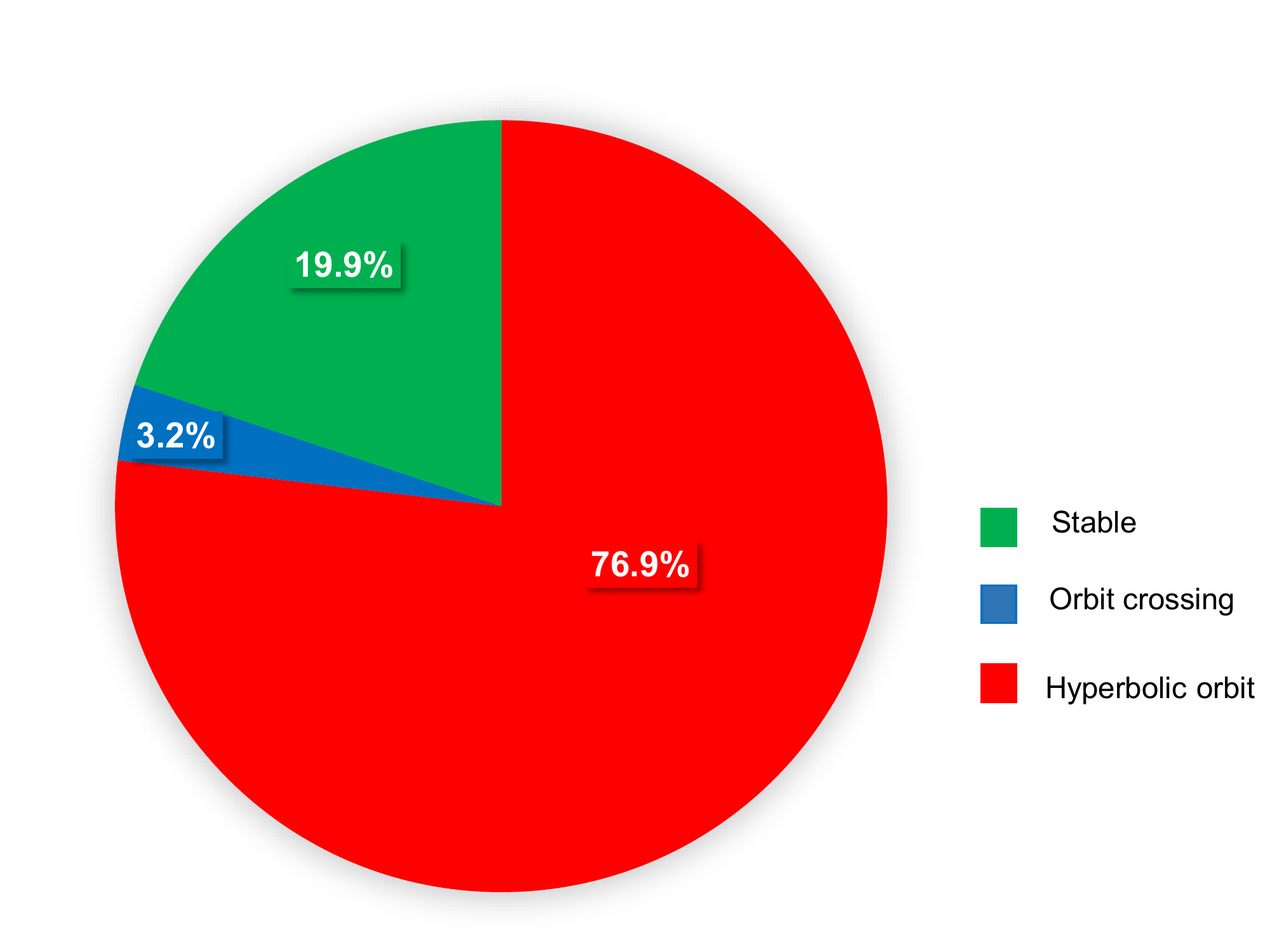}\\
\includegraphics[width=90mm,height=70mm]{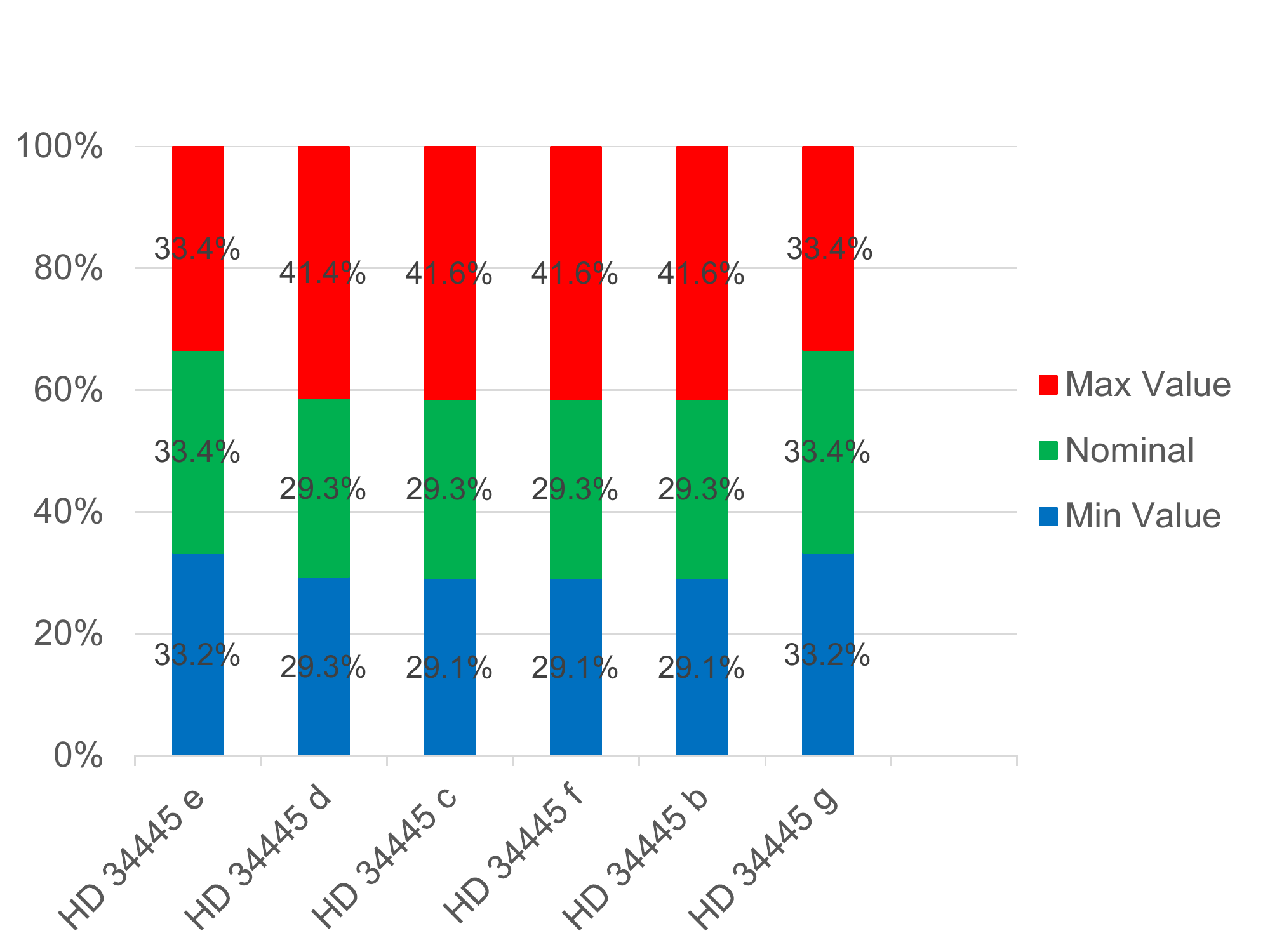}
\includegraphics[width=70mm,height=60mm]{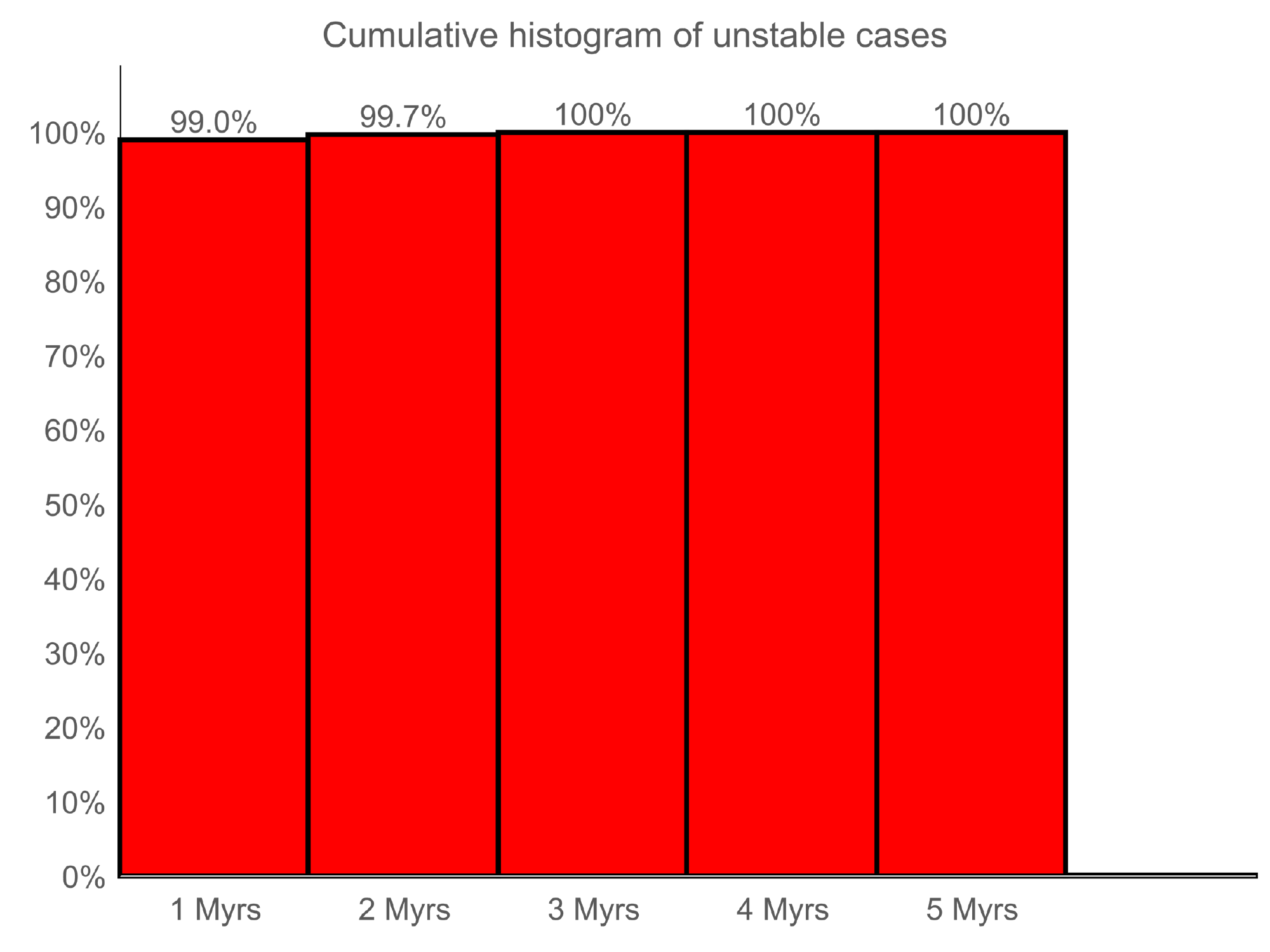}
\caption{The $99\%$ credibility interval eccentricity case integrated for five million years. Top left: percentage of hyperbolic orbits for each planet. Top right:  percentage of  planets that became hyperbolic during a simulation. Middle: percentage of stable systems, hyperbolic orbits and orbit crossing incidents. Bottom left: distribution of which of the three values (minimum, nominal, maximum) used to sample our parameters was involved in the unstable systems recorded (hyperbolic + orbit crossing cases). Bottom right: cumulative histogram of the percentage of unstable
systems with respect to the time of integration. We recorded 561 hyperbolic orbit and 23 orbit crossing cases.
\label{fig5}}
\end{center}
\end{figure*}

As one would expect, in the $99\%$ credibility interval case the number of hyperbolic orbit incidents increased dramatically. From the 135 hyperbolic orbit cases and 6 orbit crossing incidents we had in the $1\sigma$ eccentricity case, 
those numbers rose to 561 and 23 respectively in the $99\%$ credibility interval case. Regarding which planets and how many planets were involved in those cases, the situation was qualitatively similar to the $1\sigma$ eccentricity case.
The value distribution for the unstable systems was more balanced for all planets, i.d. we did not have any value that was much more dominant than the other one (as for example we had for planet HD 34445 f in the $1\sigma$ case). Finally, $99\%$ of the unstable cases were recorded within only one million years.
A graphical representation of the results of the $99\%$ credibility interval case can be found in Figure \ref{fig5}.

\subsection{Variation of the orbital inclination $i$}
 The results for the stability of the HD 34445 system while varying the orbital inclination with respect to the plane of the sky with a step of $1^{o}$showed that our planetary system is stable for $i=89^{o}$ to $i=23^{o}$.  Then from 
$i=22^{o}$ to $i=1^{o}$ the system was unstable.  The results of this numerical experiment tell us that there is a wide range of values for the inclination with respect to the plane of the sky for which the system
is dynamically stable.   Since we find unstable systems for $i=22^{o}$ or less, and $1/\sin{23^{o}}\sim 2.56$, this means that all the masses of the HD 34445 system have to increase by a factor of more that 2.5 times  before we start having our system disrupted.  The results also explain why we did not record any instability while we varied the minimum mass as part of our second numerical experiment: the mass ranges considered were well below that 2.56 factor. For instance, let us consider the case of the HD 34445 g planet whose orbital elements and mass have the greatest uncertainties among the planets of the system. Even the right endpoint of the $99\%$ credibility interval of value 257.8 divided by the nominal value of 120.6 gives a ratio of $\sim 2.13$, well below the 2.56 value we mentioned earlier.

For all the above experiments, we have monitored the changes in the energy and angular momentum of the system.  The accumulated error in those two
quantities is typically of the order of $10^{-10}$ and $10^{-12}$ respectively, although many times the errors were even smaller than those values.

\section{Summary and discussion}

In this work, we have investigated whether the RV
measurements and data fitting obtained for the planetary system HD 34445 can provide a realistic orbital solution.  The system involves  six planets, and assuming  that all bodies lie on the same plane, there were four orbital elements and the masses that we had to consider.
Ideally, we would like to vary all parameters at the same time and examine the
stability outcome of all possible combinations .  That would lead, however,  to a huge number of cases. For example, if we use three values for each parameter, then the number of all possible combinations is of the order of  $10^{14}$. Hence, it would be impossible to run the simulations in a reasonable amount of time. Therefore we chose to execute the numerical experiments described in section \ref{met}.

The nominal solution was integrated for five million years and we 
found no sign of instability during that period.  Next, we varied only one of the parameters at a time and we found that the eccentricity variation was the factor that affected the stability of the system the most.  This is to be expected as larger eccentricities mean larger probability for closer encounters among the planets 
leading eventually to orbit crossing or hyperbolic orbits.   The variation of the other parameters produced not a single unstable system for the integration time used. We thought that, possibly, the variation of the semi-major axis
may make a difference, because, as seen in Table  2, many of the planets are potentially near mean motion resonances and therefore changes in the planetary semi-major axes could lead to a more vibrant evolution of HD 34445.   For instance,
the nominal period ratio of planets HD 34445 c and HD 34445 f is 3.15. Within the $1\sigma$ value range that ratio can drop   
to 3.06.  Although we get closer to the exact resonance location, the small planetary eccentricities of the nominal solution
will probably protect the system from overlapping of neighbouring mean motion resonances that lead to chaotic motion and 
eventually, in most of the times, to escape of bodies \citep[e.g. see][for more details]{2008IAUS..246..199M,2008LNP...760...59M}.

\begin{table*}
\label{t2}
\caption[]{Period ratios between the planets  of the HD 34445 system. Each entry provides the period ratio between the planets listed horizontally over the ones listed vertically.} 
\begin{scriptsize}
\vspace{0.1 cm}
\begin{center}	
{\footnotesize \begin{tabular}{c c c c c c c c}\hline
$\backslash$ & HD 34445 g & HD 34445 b & HD 34445 f & HD 34445 c & HD 34445 d & HD 34445 e \\
\hline
HD 34445 g & 1 & 5.39 & 8.42 & 26.55 & 48.36 & 115.91&\\ 
HD 34445 b & - & 1 & 1.56 & 4.92 & 8.96 & 21.49&\\ 
HD 34445 f & - & - & 1 & 3.15 & 5.74 & 13.76&\\ 
HD 34445 c & - & - & - & 1 & 1.82 & 4.36&\\
HD 34445 d & - & - & - & - & 1 & 2.4&\\
HD 34445 e & - & - & - & - & - & 1&\\
\hline
\end{tabular}}
\end{center}
\end{scriptsize}
\end{table*}

The variation of the minimum masses was also significant in some cases. For instance, the mass variation for HD 34445 g was around $32\%$, at the $1 \sigma$ level and it grew more than $100\%$ at the $99\%$ credibility interval level.  That mass variation, however, was also not adequate to cause dynamical instability to our system. 
This was better understood when  we carried out our third numerical experiment where we found that the system was stable for a wide range of the orbital inclination $i$.  Again, it seems that the very small eccentricities of the nominal solution  decided the fate of those systems.
The findings of the third numerical experiment also support the radial velocity data in the sense that 
the wide range of stable values we have found reinforce the plausibility of the orbital solution
that has been announced.  Thanks to the wide range of masses for which we have a stable system the error 
margin we have when determining the orbital parameters of the system is larger.

Finally, our numerical experiments have assumed that all HD 34445 planets lie on the same 
orbital plane. It is likely that the orbital planes  of the planets are not perfectly aligned. It may
also be that some orbital plane is inclined more than just a few degrees. However, that may not 
affect the stability of the system as seen elsewhere \citep[e.g.][]{2013NewA...23...41G,2018MNRAS.481.2180V}. 

An interesting matter of discussion is the potential habitability of the system.  Two of the planets of our system, HD 34445 f and HD 34445 b, reside inside the habitable zone of the star, i.e. the area around
the star where a planet can have liquid water on its surface.  Considering that the effective temperature of HD 34445 is $T_{eff}=5836 K$ and using \cite{2014ApJ...787L..29K}, the habitable zone stretches from 1.330 au to 2.350 au.  These limits refer to runaway greenhouse and maximum greenhouse for a planet of mass $m_p=1 M_{\odot}$.

According to Table 1, those two planets have much larger masses than of a super Earth, even when 
we consider value fluctuations at the $99\%$ credibility interval  The minimum mass for HD 34445 f for example is just above 18 $M_{\odot}$,
i.e. slightly larger than Neptune.    As the mass estimates in Table 1 refer to minimum masses, it is very likely that the real masses of our system are larger. Our numerical simulations showed that they can even increase by a factor of more than 2.5 before we lose stability of our planetary system.  Therefore the only way to have a terrestrial sized planet in the habitable zone is
the existence of an undetected so far additional planet in the habitable zone. This, however,
may prove difficult  as the perturbations from planets HD34445 f and HD 34445 b could be significant and leave no area for a stable orbit to exist.  There is a possibility of having a wider area for the habitable zone if we consider the recent Venus and early Mars borders at
1.050 au and 2.470 au respectively. These are independent of the mass of the terrestrial planet.
On the other hand the runaway greenhouse location depends on the mass of the planet. 
 If we assume a larger 
or a smaller planet, then the runaway greenhouse border shifts inwards  (1.290 au) or outwards (1.410 au) respectively.  The maximum greenhouse border remains the same, independent of the mass of the terrestrial planet  \citep{2014ApJ...787L..29K}.  Finally, it is possible to have a slight
broadening of the habitable zone  if we consider dynamically informed habitable zones \citep [e.g.][]{2018ApJ...856..155G}. But even considering all those habitable zone stretching possibilities it may prove dynamically difficult to accommodate another planet in that region.

An alternative to having a terrestrial planet in the habitable zone of HD 34445 is  either to have an exomoon orbiting one of those two planets \citep[e.g.][]{2014AsBio..14..798H} or having a Trojan planet \citep[e.g.][]{2009CeMDA.104...69S,2012MNRAS.423.3074F}.  The fact that our planetary masses can be significantly larger implies they can host large exomoons or Trojan planets. All the above possibilities for habitable terrestrial bodies require further and in depth
investigation, especially from the gravitational point of view.

To conclude, at the $1 \sigma$ level, the HD 34445 system appears to be fairly stable. A small percentage of the total number of cases we investigated yields an unstable configuration implying that the 
discovery that was announced is plausible. 
Further observations of the system will be welcome and are necessary as the outermost planet has 
a long period and we need more data in order to reduce the uncertainties of its orbit.

\section*{Acknowledgements}
We would like to thank Siegfried Eggl who provided the code that was used for our numerical experiments. We would also like to thank the High Performance Computing Resources team at New York University Abu Dhabi and especially Jorge Naranjo for helping us with setting up our numerical simulations. Finally, we would like to thank the anonymous referee whose comments helped us improve the manuscript. 

 

\bibliographystyle{mnras}

\bsp	
\label{lastpage}
\end{document}